\documentclass[12pt,english,floatfix,superscriptaddress,aps,prd,preprint]{revtex4}
\usepackage[utf8]{inputenc}
\usepackage{amsmath}
\usepackage{amssymb}
\usepackage{amsbsy}
\usepackage{amsfonts}
\usepackage{amsopn}
\usepackage{amstext}
\usepackage{graphicx}
\usepackage{amssymb}
\usepackage{amsfonts}
\usepackage{amsmath}
\usepackage{amsmath,amsthm,amsfonts,amssymb}
\usepackage[mathcal]{eucal}
\usepackage{mathrsfs}
\usepackage{graphicx}
\usepackage[english]{babel}
\usepackage{color}
\usepackage{esint}
\usepackage[dvips]{epsfig}
\usepackage[dvips]{graphicx}
\usepackage{float}
\usepackage{units}
\usepackage{textcomp}

\usepackage{hyperref}
\usepackage{slashed}

\newcommand{\ie}{\begin{equation}}
\newcommand{\fe}{\end{equation}}
\newcommand{\se}{\begin{eqnarray}}
\newcommand{\ff}{\end{eqnarray}}

\begin{document}

\title{Exact solutions of the Dirac oscillator under the influence of the Aharonov-Casher effect in the cosmic string background}

%%%%%%%%%%%%%%%%%%%%%%%%%%%%%%%%%%%%%%%%%%%%%%%%%%%%%%%%%%%%%%%%%%%%%%

\author{R. R. S. Oliveira}
\email{rubensrso@fisica.ufc.br}
\affiliation{Universidade Federal do Cear\'a (UFC), Departamento de F\'isica,\\ Campus do Pici, Fortaleza - CE, C.P. 6030, 60455-760 - Brazil.}

%%%%%%%%%%%%%%%%%%%%%%%%%%%%%%%%%%%%%%%%%%%%%%%%%%%%%%%%%%%%%%%%%%%%%%%%

\author{R. V. Maluf}
\email{r.v.maluf@fisica.ufc.br}
\affiliation{Universidade Federal do Cear\'a (UFC), Departamento de F\'isica,\\ Campus do Pici, Fortaleza - CE, C.P. 6030, 60455-760 - Brazil.}

%%%%%%%%%%%%%%%%%%%%%%%%%%%%%%%%%%%%%%%%%%%%%%%%%%%%%%%%%%%%%%%%%%%%%%

\author{C. A. S. Almeida}
\email{carlos@fisica.ufc.br}
\affiliation{Universidade Federal do Cear\'a (UFC), Departamento de F\'isica,\\ Campus do Pici, Fortaleza - CE, C.P. 6030, 60455-760 - Brazil.}

\date{\today}

\begin{abstract}

In this work, we study the Aharonov-Casher effect in the $(2+1)$-dimensional Dirac oscillator coupled to an external electromagnetic field. We set up our system in two different scenarios: in the Minkowski spacetime and the cosmic string spacetime. In both cases, we solve exactly the Dirac oscillator and we determine the energy spectrum and the eigenfunctions for the bound states. We verify that in the Minkowski spacetime, the Dirac oscillator spectrum depends linearly on the strength of the magnetic field $B$, and on the Aharonov-Casher phase. In addition, we explicitly obtain the corrections on the Dirac spinors and the energy levels due to the curvature effects in the cosmic string background. Finally, we investigate the nonrelativistic limit and compare our results with those found in the literature.

\end{abstract}

%\keywords{*****}
\maketitle

%%%%%%%%%%%%%%%%%%%%%%%%%%%%%%%%%%%%%%%%%%%%%%%%%%%%%%%%%%%%%%%%%%%%%%%%%%%%%%%%%%

\section{Introduction}

Since it was introduced in the literature by Moshinsky and Szczepaniak \cite{Moshinsky}, the Dirac oscillator (DO) has been study in several context, such as in thermal physics \citep{Pacheco}, graphene system \cite{Quimbay2013cond-mat.}, quantum phase transitions \cite{Bermudez}, noncommutative space \cite{Mandal}, cosmic string spacetime \cite{Andrade2014}, minimal length \cite{Quesne}, and etc. To set up the DO we apply a nonminimal substitution in the Dirac equation (DE) given by ${\bf p}\rightarrow {\bf p}-im_{0}\omega \beta {\bf r}$, where $m_{0}$ is the rest mass of the fermion, $\omega>0$ is the angular frequency of the DO and {\bf{r}} is the position vector \cite{Moshinsky}. Recently, the DO has been verified experimentally by J. A. Franco-Villafa\~ne $et$ $al$. \cite{Franco}.

On the other hand, the study of topological quantum phases has also attracted a great deal of attention in recent years. An interesting example of this study was carried out by Aharonov and Casher \cite{Aharonov}, who verified the appearance of a topological phase in the wave function of a neutral particle with magnetic dipole moment (MDM) that interacts with an external electric field \cite{He}. This effect came to be known as the Aharonov-Casher (AC) effect. Subsequently, it was proposed a relativistic treatment for the AC effect, in particular, for spin-$1/2$ neutral fermions  \cite{Hagen1990,Silva,Li}. Besides, the formalism of the DE for neutral fermions with MDM in the presence of electromagnetic fields have various physics applications, for example, it is applicable in problems of scattering \cite{Lin}, spin effects \cite{Azevedo1}, in the presence of gravity and non-inertial frames \cite{Bakke2010}, topological defects \cite{Bakke2008}, quantum dot \cite{Bakke2012}, etc.  

Also, some attention has been given to the DE in curved spacetime \cite{Carter,Boada,Cariglia,Li2008}, or, also under the influence of curvature due to topological defects \cite{Medeiros,Belich}. It has been suggested that relativistic quantum systems in a curved space show interesting effects due to gravity  \cite{Obukhov2001}. Direct evidence of this is based on neutron interferometry technique \cite{Colella, Stodolsky,Cai}, in which it has shown that even on the Earth's surface or nearby space, gravitational effects in quantum systems are measurable.

The present work has as its main goal to investigate the relativistic quantum dynamics of the $(2+1)$-dimensional DO in the presence of the AC effect coupled to an external electromagnetic field in two different scenarios: in the Minkowski spacetime and the cosmic string spacetime. In the literature, one of the first paper that studied the dynamics of the DO in the presence of the AC effect was carried out by Bakke and Furtado \cite{Bakke2013}. These authors studied the $(3+1)$-dimensional DO in different scenarios of general relativity. Contrary to the work of Bakke and Furtado, we assume here that the dynamics of the DO is purely planar and subjected to the presence of an external electromagnetic field. The planar solutions of the DE are not only interesting in itself but also allows connection to the rich physics of two-dimensional fermions in condensed matter systems, such as graphene \cite{Quimbay2013cond-mat.,Jellal,Novoselov}, topological insulators \cite{Kane2} and electron gas \cite{Novoselov2}.

In the AC effect, the configuration of the electric field ${\bf E}_1$ is generated by an infinitely long wire and uniformly charged located along on the $z$-axis perpendicular to the polar plane. Explicitly, this electric field is given by \cite{Aharonov,Hagen1990,Silva,Li}
\ie {\bf E}_1=\frac{2\lambda_1}{\rho}\hat{e}_{\rho}, \ \  \nabla\cdot{\bf E}_1=2\lambda_1\frac{\delta(\rho)}{\rho}, \ \ (\rho=\sqrt{x^2+y^2}), 
\label{field1}\fe
where $\lambda_1=\lambda_{0}/4\pi\epsilon_0$, being $\lambda_{0}>0$ the electric charge linear density of the wire and $\rho>0$ is the radial coordinate. With respect to the external electromagnetic field, we consider a homogeneous magnetic field ${\bf B}$ originated by an infinitely long solenoid and an electric field ${\bf E}_2$ in the inner region of a uniformly charged non-conducting cylinder of length $L$ and radius $R$. Therefore, this electromagnetic field (cross fields) is written as follows
\ie {\bf E}_2=\frac{\lambda_2\rho}{2}\hat{e}_{\rho}, \ \  \nabla\cdot{\bf E}_2=\lambda_2, \ \ \frac{\partial{\bf E}_2}{\partial t}=0, \ \ \nabla\times{\bf E}_2=0, 
\label{field2}\fe
\ie {\bf B}=B\hat{e}_{z}, \ \  \nabla\cdot{\bf B}=0, \ \ \frac{\partial{\bf B}}{\partial t}=0, \ \ \nabla\times{\bf B}=\mu_0{\bf J}, 
\label{field3}\fe
where $\lambda_2=\chi/\epsilon_0$, being $\chi=(\frac{Q}{\pi R^2L})>0$ ($R\ll L$), the electric charge volumetric density of cylinder and $B$ the strength of magnetic field. The special configuration of fields \eqref{field2} and \eqref{field3} was proposed initially by Ericsson and Sj\"oqvist to study an atomic analog of the Landau quantization based on the AC effect \cite{Ericsson}. It is worth mentioning that this electromagnetic field applied to spin-$1/2$ neutral fermions with MDM (or electric dipole moment) is of interest in relativistic \cite{Bakke2010,Bakke2012,Bakke} and nonrelativistic quantum systems \cite{Ribeiro2006}.

This work is organized as follows. In Section \ref{sec2}, we introduce the DO in the Minkowski spacetime in the presence of a electromagnetic field and of the AC effect, just as we determine the energy spectrum and eigenfunctions to the bound-states of the system. In Section \ref{sec3}, we study the influence of curvature of the cosmic string spacetime in the DO in the presence of a electromagnetic field and of the AC effect. In Section \ref{sec4}, we take the nonrelativistic limit of our results. In Section \ref{conclusion}, we present our conclusions.

%-------------------------------------------------------------------
\section{The Dirac Oscillator and the  Aharonov-Casher effect in the Minkowski spacetime\label{sec2}}

In this section, we study the DO in the presence of the AC effect with an external electromagnetic field in the Minkowski spacetime. We consider the polar coordinate system $(t,\rho,\theta)$ whose signature of the metric tensor is $(+,-,-)$. Thus, the Minkowski spacetime is described by the following line element
\ie ds^2=c^2dt^2-d\rho^2-\rho^2 d\theta^2,
\label{lineelement}\fe
where $-\infty<t<\infty$, ${0}\leq\theta\leq{2\pi}$ is the azimutal coordinate and $c$ is the speed of light \cite{Greiner}. 

The covariant DE that governs the dynamics of a neutral fermion with MDM $\mu$ is written via nonminimal coupling as follows (in natural units $\hbar=c=1$) \cite{Greiner}
\ie \left[i\gamma^{a}\partial_{a}+\frac{\mu}{2}\sigma^{ab}F_{ab}-m_{0}\right]\Psi(t,{\bf r})=0, \ \ (a,b=0,1,2),
\label{Dirac1}\fe
where $\gamma^{a}$ are the gamma matrices and satisfies the anticommutation relation of Clifford Algebra: $\{\gamma^{a},\gamma^{b}\}=2\eta^{ab}\mathbb{I}_{2\times 2}$, being $\eta^{ab}$ the Minkowski metric tensor, $\sigma^{ab}=\frac{i}{2}[\gamma^{a},\gamma^{b}]$, $F_{ab}$ is the electromagnetic field tensor and $\Psi(t,{\bf r})$ is the two-component Dirac spinor, respectively. As we shall see in the following section, the indices given by the Greek letters ($\mu, \nu, \sigma,\ldots$) refer to curved spacetime and those of the Latin letter to the flat spacetime ($a, b, c, \ldots$). By substituting the nonminimal coupling of the DO and using the relation $\frac{1}{2}\sigma^{ab}F_{ab}=i\boldsymbol{\alpha}\cdot{\bf E}-\boldsymbol{\Sigma}\cdot{\bf B}$, where {\bf E} and {\bf B} are the electric and magnetic fields, the Eq. \eqref{Dirac1} can be written as 
\ie \left[i\gamma^{0}\partial_{0}-\boldsymbol{\gamma}\cdot({\bf p}-im_0\omega\beta{\bf r})+\mu(i\boldsymbol{\alpha}\cdot{\bf E}-\boldsymbol{\Sigma}\cdot{\bf B})-m_{0}\right]\Psi(t,{\bf r})=0.
\label{Dirac2}\fe

Replacing now the fields \eqref{field1}, \eqref{field2} and \eqref{field3} into Eq. \eqref{Dirac2}, we get
\ie \left[i\gamma^{0}\partial_0+i\gamma^{\rho}\left(\partial_\rho+\gamma^{0}\left(m_{0}\bar{\omega}\rho-\frac{s\Phi_{AC}}{\pi\rho}\right)\right)+\frac{i\gamma^{\theta}\partial_\theta}{\rho}-\Sigma^{z}\mu B-m_{0}\right]\Psi(t,\rho,\theta)=0,
\label{Dirac3}\fe
where 
\ie \gamma^0=\beta, \ \gamma^{\rho}=\boldsymbol{\gamma}\cdot\hat{e}_{\rho}=\gamma^{1}\cos\theta+\gamma^{2}\sin\theta, \ \gamma^{\theta}=\boldsymbol{\gamma}\cdot\hat{e}_{\theta}=-\gamma^{1}\sin\theta+\gamma^{2}\cos\theta, \ \Sigma^{z}=\boldsymbol{\Sigma}\cdot\hat{e}_{z},
\label{matrices}\fe
being $\bar{\omega}=(\omega-\frac{\omega_{AC}}{2})\geqslant{0}$, where $\omega_{AC}=\frac{\mu\lambda_2}{m_{0}}$ is the Aharonov-Casher (AC) frequency of the fermion \cite{Ericsson} and $\Phi_{AC}=2s\pi\mu\lambda_1$ is the Aharonov-Casher (AC) phase, where $s=\pm 1$ corresponds to the projections of the MDM of the fermion along on the $z$-axis \cite{He,Bakke2013}. Besides that, through a similarity transformation given by unitary operator $U(\theta)=e^{-\frac{i\theta\alpha_3}{2}}$ \cite{Villalba}, we can reduce the Dirac matrices $\gamma^{\rho}$ and $\gamma^{\theta}$ to fixed matrices $\gamma^{1}$ and $\gamma^{2}$ as follows
\ie U^{-1}(\theta)\gamma^{\rho}U(\theta)=\gamma^{1}, \ \  U^{-1}(\theta)\gamma^{\theta}U(\theta)=\gamma^{2}.
\label{unitary}\fe

Since we are working in a $(2+1)$-dimensional Minkowski spacetime, it is convenient to define the Dirac matrices $\boldsymbol{\gamma}=(\gamma^1,\gamma^2)=(-\gamma_1,-\gamma_2)$, $\gamma^{0}$ and $\Sigma^z$ in terms of the Pauli matrices, i.e., $\gamma_1=\sigma_3\sigma_1$, $\gamma_2=\sigma_3\sigma_2$ and $\gamma^{0}=\Sigma^z=\sigma_3$ \citep{Greiner,Villalba}. So, by using this information and the relations \eqref{unitary}, we transform the Eq. \eqref{Dirac3} into the form:
\ie \left[\sigma_{1}\left(i\partial_\rho+i\sigma_{3}\left(m_{0}\bar{\omega}\rho-\frac{s\Phi_{AC}}{\pi\rho}\right)\right)+\sigma_{2}\left(\frac{i\partial_\theta}{\rho}+\frac{\sigma_3}{2\rho}\right)+\mu B+m_{0}\sigma_{3}-i\partial_0\right]\psi(t,\rho,\theta)=0,
\label{Dirac4}\fe
where $\psi(t,\rho,\theta)\equiv U^{-1}(\theta)\Psi(t,\rho,\theta)$. 

Assuming the following ansatz for the Dirac spinor \cite{Villalba,Andrade}
\ie \psi(t,\rho,\theta)=\frac{e^{i(m_{l}\theta-Et)}}{\sqrt{2\pi}}\left(
           \begin{array}{c}
            \phi_{+}(\rho) \\
             i\phi_{-}(\rho) \\
           \end{array}
         \right), \ \ (m_{l}=\pm1/2,\pm3/2,\ldots)
\label{spinor},\fe we obtain from \eqref{Dirac4}, a system of two first-order coupled differential equations given by
\ie (m_0+\mu B-E)\phi_+(\rho)=\left[\frac{d}{d\rho}-m_0\bar{\omega}\rho+\frac{1}{\rho}\left(m_l+\frac{s\Phi_{AC}}{\pi}+\frac{1}{2}\right)\right]\phi_-(\rho)
\label{Dirac5}, \fe
\ie (m_0-\mu B+E)\phi_-(\rho)=\left[\frac{d}{d\rho}+m_0\bar{\omega}\rho-\frac{1}{\rho}\left(m_l+\frac{s\Phi_{AC}}{\pi}-\frac{1}{2}\right)\right]\phi_+(\rho)
\label{Dirac6}. \fe

Substituting \eqref{Dirac6} into \eqref{Dirac5} and vice versa, we get two differential equations written compactly as 
\ie \left[\frac{d^{2}}{d\rho^{2}}+\frac{1}{\rho}\frac{d}{d\rho}-\frac{\gamma_s^{2}}{\rho^{2}}-m_{o}^{2}\bar{\omega}^{2}\rho^{2}+E_{s}\right]\phi_{s}(\rho)=0, \ \ (s=\pm 1)
\label{Dirac7}, \fe
where
\ie \gamma_s\equiv m_{l}+\frac{s\Phi_{AC}}{\pi}-\frac{s}{2}, \ \  E_{s}\equiv(\mu B-E)^{2}-m_{0}^{2}+2m_{0}\bar{\omega}\gamma_s+2sm_{0}\bar{\omega}
\label{Dirac8}, \fe
being $\phi_{s}(\rho)$ real radial functions, $E$ is the relativistic total energy of the fermion and $m_{l}$ is the orbital magnetic quantum number. Besides that, the connection between $m_l$ and $m_j$, where $m_j$ is the total magnetic quantum number, it is given as follows
\ie J_z\psi(t,\rho,\theta)=(m_l+m_s)\psi(t,\rho,\theta)=m_j\psi(t,\rho,\theta)
\label{Dirac9}, \fe
where $J_z=L_z+S_z$ is the projection of the total angular momentum operator ${\bf J}$ in the z-axis, being $L_z=-i\frac{\partial}{\partial\theta}$ and $S_z=\frac{1}{2}\sigma_3$, $m_s=\frac{s}{2}$ is the spin magnetic quantum number and the values possibles of $m_j$ are $m_j=0,\pm 1, \pm 2,\ldots$ \cite{Villalba,Andrade}.

In order to solve the Eq. \eqref{Dirac7}, we will introduce a new variable given by $\tau=m_{0}\vert\bar{\omega}\vert\rho^2$. Thereby, the Eq. \eqref{Dirac7} becomes
\ie \left[\tau\frac{d^{2}}{d\tau^{2}}+\frac{d}{d\tau}-\frac{\gamma_s^{2}}{4\tau}-\frac{\tau}{4}+\frac{E_{s}}{4m_{0}\vert\bar{\omega}\vert}\right]\phi_{s}(\tau)=0
\label{Dirac10}.\fe

Analyzing the asymptotic behavior of the Eq. (\ref{Dirac10}) for $\tau\to{0}$ and $\tau\to{\infty}$, we can write a regular solution at the origin in the form
\ie \phi_{s}(\tau)=C_{s}\tau^{\frac{\vert\gamma_s\vert}{2}}e^{-\frac{\tau}{2}}R_{s}(\tau)
\label{Dirac11},\fe 
where $C_{s}$ are normalization constants and $R_{s}(\tau)$ are unknown functions to be determined.

In this way, substituting \eqref{Dirac11} into \eqref{Dirac10}, we obtain
\ie \left[\tau\frac{d^{2}}{d\tau^{2}}+(\vert \gamma_s\vert+1-\tau)\frac{d}{d\tau}+\left(\frac{E_{s}}{4m_{0}\vert\bar{\omega}\vert}-\frac{\vert \gamma_s\vert+1}{2}\right)\right]R_{s}(\tau)=0
\label{Dirac12}.\fe

It is not difficult to note that Eq. \eqref{Dirac12}  takes the form of a confluent hypergeometric equation, whose solution is the confluent hypergeometric function $R_{s}(\tau)=\mathstrut_1 F_1\left(\frac{\vert \gamma_s\vert+1}{2}-\frac{E_{s}}{4m_{0}\vert\bar{\omega}\vert}, \vert \gamma_s\vert+1, \tau\right)$ \cite{Abramowitz}. For normalizable solutions, the confluent hypergeometric function must become a polynomial of degree $n$ \cite{Abramowitz}, then, we must impose the parameter $\frac{\vert \gamma_s\vert+1}{2}-\frac{E_{s}}{4m_{0}\vert\bar{\omega}\vert}$ to be equal to a non-positive integer number $-n$ ($n=0,1,2,\ldots$). Therefore, using this imposition and the relation \eqref{Dirac8}, we obtain the following energy spectrum of the DO in the Minkowski spacetime under the influence of an electromagnetic field and the AC effect
\ie E_{n_s,m_l}=\mu B\pm\sqrt{m_{0}^{2}+4m_{0}{\vert\bar{\omega}\vert}\left[n_s+\frac{\vert \gamma_s \vert}{2}-\frac{\gamma_s}{2}\right]},
\label{spectrum}\fe
where $n_s=n+\frac{1-s}{2}$ is a quantum number, the $+$ sign corresponds to the positive energies states for the particle and the $-$ sign corresponds to the negative energies states for the antiparticle, respectively. We can see that the energy spectrum \eqref{spectrum}, besides of depending linearly on the strength of the magnetic field $B$, also depends on the Aharonov-Casher frequency $\omega_{AC}$ and on the  Aharonov-Casher phase $\Phi_{AC}$ with periodicity $\Phi_0=\pm 2\pi$, where we have $E_{n_s,m_l}(\Phi_{AC}\pm 2\pi)=E_{n_s,m_l+1}(\Phi_{AC})$. We observed still that in the case where the DO is tuned to resonate perfectly with the half-frequency of $\omega_{AC}$ ($\vert\bar{\omega}\vert=0$), the total energy of the particle and antiparticle is a physics constant that depends on its mass of rest and of the energy acquired by the magnetic field.

Comparing the energy spectrum \eqref{spectrum} with the results of the literature, we see that in absence of the external electromagnetic field ($B=\omega_{AC}=0$) with $s=+1$, this spectrum is reduced to the energy spectrum of the Ref. \cite{Bakke2013} (with $k=cte=0$). Last but not least, also we verified that for $\mu=0$, the spectrum \eqref{spectrum} is reduced to the energy spectrum of DO usual \cite{Villalba,Andrade}.

Now, let us concentrate on the form of the original Dirac spinor to bound states of the system. Initially, let us obtain the form of the rotated Dirac spinor \eqref{spinor}. So, substituting the variable $\tau=m_0\vert\bar{\omega}\vert\rho^2$ in the radial function \eqref{Dirac11} and posteriorly substituting this function in the spinor \eqref{spinor}, we obtain
\ie \psi_{n,m_l}(t,\rho,\theta)=e^{i(m_{l}\theta-Et)}\left(
           \begin{array}{c}
            \bar{C}_{+}\rho^{\vert\gamma_{+}\vert}e^{-\frac{m_0\vert\bar{\omega}\vert\rho^2}{2}}\mathstrut_1 F_1\left(-n, \vert \gamma_{+}\vert+1, m_0\vert\bar{\omega}\vert\rho^2\right) \\
             i\bar{C}_{-}\rho^{\vert\gamma_{-}\vert}e^{-\frac{m_0\vert\bar{\omega}\vert\rho^2}{2}}\mathstrut_1 F_1\left(-n, \vert \gamma_{-}\vert+1, m_0\vert\bar{\omega}\vert\rho^2\right) \\
           \end{array}
         \right)
\label{spinor2},\fe
where we define a new normalization constant given by
\ie \bar{C}_{s}\equiv\frac{C_s(m_0\vert\bar{\omega}\vert)^{\frac{\vert\gamma_s\vert}{2}}}{\sqrt{2\pi}}
\label{constant}.\fe

Writing the original Dirac spinor as $\Psi(t,\rho,\theta)=e^{-\frac{i\theta\sigma_3}{2}}\psi(t,\rho,\theta)$, we have
\ie \Psi_{n,m_l}(t,\rho,\theta)=e^{-iEt}\left(
           \begin{array}{c}
            \bar{C}_{+}\rho^{\vert\gamma_{+}\vert}e^{i\left(m_{l}-\frac{1}{2}\right)\theta}e^{-\frac{m_0\vert\bar{\omega}\vert\rho^2}{2}}\mathstrut_1 F_1\left(-n, \vert \gamma_{+}\vert+1, m_0\vert\bar{\omega}\vert\rho^2\right) \\
             i\bar{C}_{-}\rho^{\vert\gamma_{-}\vert}e^{i\left(m_{l}+\frac{1}{2}\right)\theta}e^{-\frac{m_0\vert\bar{\omega}\vert\rho^2}{2}}\mathstrut_1 F_1\left(-n, \vert \gamma_{-}\vert+1, m_0\vert\bar{\omega}\vert\rho^2\right) \\
           \end{array}
         \right)
\label{spinor3}.\fe

It should be noted that our result incorporates the positive and negative values of the quantum number $m_l$, which does not happen for the case of the DO worked in Refs. \cite{Villalba,Andrade}.

%-------------------------------------------------------------------
\section{The Dirac Oscillator and the  Aharonov-Casher effect in the cosmic string spacetime\label{sec3}}

Now, we will consider the previously studied system in the presence of a curved background. The curvature effects are due to a topological defect background, given by the cosmic string spacetime. The influence of topological defects on physical systems has been studied in several works \cite{Vitoria,Gerbert,Marques,Dantas,Bausch}, and the cosmic string is a simple example of a defect that includes curvature and can be applied in various contexts \cite{Bakke2013,Medeiros,Belich}.

Following a usual procedure \cite{Carvalho,Bakke2013}, we start from the DO in cylindrical coordinates; posteriorly, we turn the system into a dynamics purely planar. Thus, the cosmic string spacetime is described by the following line element \cite{Vilenkin,Vachaspati,Hindmarsh} 
\ie ds^2=c^2dt^2-d\rho^2-\eta^2\rho^2 d\theta^2-dz^2,
\label{lineelement2}\fe
where $-\infty<(t,z)<\infty$ and $\eta=1-\frac{4\bar{m}G}{c^2}$ is a parameter related to the deficit angle, being $\bar{m}$ the linear mass density of the cosmic string  \cite{Medeiros,Bakke2013,Carvalho}. Here, the deficit angle can only assume values where $0<\eta\leq 1$, unlike assumed in \cite{Katanaev,Furtado1994}, where $\eta$ can take values greater than 1, which corresponds to an anti-conical spacetime with negative curvature. 

The geometry characterized by the line element \eqref{lineelement2}  has a  conical singularity that gives rise to the curvature centered on the cosmic string axis ($z$-axis). However, in all other places the curvature is null. This conical singularity is represented by the following curvature tensor
\ie R^{\rho,\theta}_{\rho,\theta}=\frac{1-\eta}{4\eta}\delta_2(\bf r),
\label{lineelement}\fe
where $\delta_2(\bf r)$ is the two-dimensional Dirac delta \cite{Andrade2014,Medeiros,Bakke2013}.

The covariant DO that governs the dynamics of a neutral fermion with MDM $\mu$ is written via nonminimal coupling as follows (in natural units $\hbar=c=G=1$) \cite{Bakke2013}
\ie \left[i\gamma^{\mu}(x)\left(\nabla_{\mu}(x)+m_0\omega\rho\gamma^{0}\delta^{\rho}_\mu\right)+\frac{\mu}{2}\sigma^{\mu\nu}(x)F_{\mu\nu}-m_{0}\right]\Psi(t,{\bf r})=0, \ \ (\mu,\nu=0,1,2,3),
\label{string1}\fe
where $\gamma^{\mu}(x)$ are the gamma matrices and satisfies the anticommutation relation of covariant Clifford Algebra: $\{\gamma^{\mu}(x),\gamma^{\nu}(x)\}=2g^{\mu\nu}(x)\mathbb{I}_{4\times 4}$, being $g^{\mu\nu}(x)$ the curved metric tensor, $\nabla_{\mu}(x)=\partial_\mu+\Gamma_\mu(x)$ is the covariant derived, being $\Gamma_\mu(x)$ the spinor affine connection and $\delta^{\rho}_\mu$ is the Kronecker delta. The connection between $\gamma^{\mu}(x)$ with the gamma matrices defined in the Minkowski spacetime $\gamma^{a}$ is given by $\gamma^{\mu}(x)=e^{\mu}_a(x)\gamma^{a}$, where $e^{\mu}_a(x)$ is the basis tetrad and satisfies the conditions: $e^{\mu}_a(x)e^{\nu}_b(x)\eta^{ab}=g^{\mu\nu}(x)$, $e^{a}_\mu(x)e^{\mu}_b(x)=\delta^{a}_{b}$ and $e^{\mu}_a(x)e^{a}_\nu(x)=\delta^{\mu}_{\nu}$, being $\eta^{ab}$ the Minkowski metric tensor ($a,b=0,1,2,3$). Explicitly, the basis tetrad is defined in the cosmic string spacetime as  \cite{Andrade2014,Carvalho,Bakke2013}
\ie e^{a}_\mu(x)=\left(\begin{array}{cccc}
1 & 0 & 0 & 0	\\
0 & \cos\theta & -\eta\rho\sin\theta & 0	\\
0 & \sin\theta & \eta\rho\cos\theta & 0	\\
0 & 0 & 0 & 1
\end{array}\right), \ \ e^{\mu}_a(x)=\left(\begin{array}{cccc}
1 & 0 & 0 & 0	\\
0 & \cos\theta & \sin\theta & 0	\\
0 & -\frac{\sin\theta}{\eta\rho} & \frac{\cos\theta}{\eta\rho} & 0	\\
0 & 0 & 0 & 1
\end{array}\right).
\label{tetrad} \fe

Consequently, this basis tetrad allows us to write the matrices $\gamma^{\mu}(x)$ as follows
\ie \gamma^{0}(x)=\gamma^{t}, \ \ \gamma^{1}(x)=\gamma^{\rho}, \ \ \gamma^{2}(x)=\frac{\gamma^{\theta}}{\eta\rho}, \ \ \gamma^{3}(x)=\gamma^{z}
\label{matrices},\fe
where $\gamma^{\rho}=\gamma^{1}\cos\theta+\gamma^{2}\sin\theta$ and $\gamma^{\theta}=-\gamma^{1}\sin\theta+\gamma^{2}\cos\theta$. 

Thereby, we rewrite the DO \eqref{string1} in terms of basis tetrad in the following form
\ie \left[ie^{\mu}_a(x)\gamma^{a}\left(\nabla_{\mu}(x)+m_0\omega\rho\gamma^{0}\delta^{\rho}_\mu\right)+\frac{\mu}{2}e^{\mu}_a(x)e^{\nu}_b(x)\sigma^{ab}F_{\mu\nu}-m_{0}\right]\Psi(t,{\bf r})=0.
\label{string2}\fe

Besides that, the spinor affine connection $\Gamma_\mu(x)$ its written as $\Gamma_\mu(x)=\frac{i}{4}\omega_{\mu ab}(x)\sigma^{ab}$, where $\omega_{\mu ab}(x)$ is the spin connection, given by
\ie \omega_{\mu ab}(x)=\eta_{ac}[e^{c}_\nu(x)e^{\sigma}_b(x)\Gamma^{\nu}_{\sigma\mu}-e^{c}_\nu(x)\partial_\mu e^{\nu}_b(x)],
\label{spin}\fe
being $\Gamma^{\nu}_{\sigma\mu}$ the Christoffel symbol of the second kind. Furthermore, by taking the tetrads \eqref{tetrad} and solving the Maurer-Cartan structure equations in the absence of torsion \cite{Andrade2014,Bakke2013}, we
obtain the following non-null components of $\omega_{\mu ab}(x)$: $\omega^{1}_{\theta 2}(x)=-\omega^{2}_{\theta 1}(x)=(1-\eta)$. So, substituting this result in the spinor affine connection $\Gamma_{\mu}(x)$, we get: $\Gamma_{\theta}(x)=\frac{i}{2}(1-\eta)\gamma^{3}$. Therefore, we have: $\gamma^{\mu}(x)\Gamma_{\mu}(x)=-\frac{(1-\eta)}{2\eta\rho}\gamma^{\rho}$, where we use the property $\gamma^{\theta}\gamma^{3}=i\gamma^{\rho}$ \cite{Carvalho}.

In this point, we are ready to configure our system in the cosmic string background. Using the matrices \eqref{matrices}, we transform the Eq. \eqref{string2} as follows
\ie \left[i\gamma^{0}\partial_0+i\gamma^{\rho}\left(\partial_\rho-\frac{(1-\eta)}{2\eta\rho}+m_{0}\bar{\omega}\rho\gamma^{0}\right)+\frac{i\gamma^{\theta}\partial_\theta}{\eta\rho}+i\gamma^{z}\partial_z+\mu(i\boldsymbol{\alpha}\cdot{\bf E}-\boldsymbol{\Sigma}\cdot{\bf B})-m_{0}\right]\Psi=0.
\label{string3}\fe

So, substituting the fields \eqref{field1}, \eqref{field2} and \eqref{field3} into Eq. \eqref{string2} and adopting the polar coordinates system $(t,\rho,\theta)$ where $p_z=z=0$, we have
\ie \left[i\gamma^{0}\partial_0+i\gamma^{\rho}\left(\partial_\rho-\frac{(1-\eta)}{2\eta\rho}+\gamma^{0}\left(m_{0}\tilde{\omega}\rho-\frac{s\Phi_{AC}}{\eta\pi\rho}\right)\right)+\frac{i\gamma^{\theta}\partial_\theta}{\eta\rho}-\frac{\Sigma^{z}\mu B}{\eta}-m_{0}\right]\Psi(t,\rho,\theta)=0,
\label{string4}\fe
being $\Sigma^{z}=\boldsymbol{\Sigma}\cdot\hat{e}_{z}$, $\tilde{\omega}=(\omega-\frac{\tilde{\omega}_{AC}}{2})\geqslant{0}$, where $\tilde{\omega}_{AC}=\frac{\mu\lambda_2}{\eta m_0}$. Here, the tilde is used to differentiate the quantities with respect to the flat case. As the fields \eqref{field1}, \eqref{field2} and \eqref{field3} are superposed to the cosmic string, i.e., coincide in the $z$-axis, we made the following modification in the electric and magnetic fields in \eqref{string4}. namely ${\bf E}\to\frac{{\bf E}}{\eta}$ and ${\bf B}\to\frac{{\bf B}}{\eta}$ \cite{Bakke2008,Bakke2013}.

Using the procedure described above in which the matrices $\gamma^{\rho}$ and $\gamma^{\theta}$ are transformed in the matrices $\gamma^{1}$ and $\gamma^{2}$, the Eq. \eqref{string4} takes the form
\ie \left[i\sigma_{1}\left(\partial_\rho-\frac{(1-\eta)}{2\eta\rho}+\sigma_{3}\left(m_{0}\tilde{\omega}\rho-\frac{s\Phi_{AC}}{\eta\pi\rho}\right)\right)+\sigma_{2}\left(\frac{i\partial_\theta}{\eta\rho}+\frac{\sigma_3}{2\eta\rho}\right)+\mu B+\sigma_{3}m_{0}-i\partial_0\right]\psi=0,
\label{string5}\fe
where $\psi(t,\rho,\theta)\equiv U^{-1}(\theta)\Psi(t,\rho,\theta)$. 

Following the same definitions used in the previous section, it can be shown that the differential equation for the radial component of the spinor $\psi$ in cosmic string space is given by
\ie \left[\frac{d^{2}}{d\rho^{2}}+\frac{1}{\rho}\frac{d}{d\rho}-\frac{\tilde{\gamma}_s^{2}}{\eta^2\rho^{2}}-m_{0}^{2}\tilde{\omega}^{2}\rho^{2}+\tilde{E}_{s}\right]\tilde{\phi}_{s}(\rho)=0
\label{string9}, \fe
where
\ie \tilde{\gamma}_s\equiv m_l+\frac{s\Phi_{AC}}{\pi}-\frac{s\eta}{2}, \ \  \tilde{E}_{s}\equiv\left(\frac{\mu B}{\eta}-E\right)^{2}-m_{0}^{2}+\frac{2m_{0}\tilde{\omega}\tilde{\gamma}_s}{\eta}+2sm_{0}\tilde{\omega}
\label{string10}.\fe

In terms of the variable $\tau=m_{0}\vert\tilde{\omega}\vert\rho^2$, the equation \eqref{string9} becomes
\ie \left[\tau\frac{d^{2}}{d\tau^{2}}+\frac{d}{d\tau}-\frac{\tilde{\gamma}_s^{2}}{4\eta^2\tau}-\frac{\tau}{4}+\frac{\tilde{E}_{s}}{4m_{0}\vert\tilde{\omega}\vert}\right]\tilde{\phi}_{s}(\tau)=0,
\label{string11}\fe and can be transformed into the form of a confluent hypergeometric equation assuming, as before, a regular solution at the origin:
\ie \tilde{\phi}_{s}(\tau)=\tilde{C}_{s}\tau^{\frac{\vert\tilde{\gamma}_s\vert}{2\eta}}e^{-\frac{\tau}{2}}\tilde{R}_{s}(\tau)
\label{string12},\fe 
where $\tilde{C}_{s}$ are normalization constants and $\tilde{R}_{s}(\tau)$ are unknown functions to be determined.

In this way, substituting \eqref{string12} into \eqref{string11}, we obtain
\ie \left[\tau\frac{d^{2}}{d\tau^{2}}+\left(\frac{\vert \tilde{\gamma}_s\vert}{\eta}+1-\tau\right)\frac{d}{d\tau}+\left(\frac{\tilde{E}_{s}}{4m_{0}\vert\tilde{\omega}\vert}-\frac{\vert\tilde{\gamma}_s\vert}{2\eta}-\frac{1}{2}\right)\right]\tilde{R}_{s}(\tau)=0
\label{string13},\fe whose solution is the confluent hypergeometric function $\tilde{R}_{s}(\tau)=\mathstrut_1 F_1\left(\frac{\vert\tilde{\gamma}_s\vert}{2\eta}+\frac{1}{2}-\frac{\tilde{E}_{s}}{4m_{0}\vert\tilde{\omega}\vert},\frac{\vert\tilde{\gamma}_s\vert}{\eta}+1, \tau\right)$ \cite{Abramowitz}. Applying again the criterion for a normalizable solution, we obtain the following energy spectrum of the DO in the cosmic string spacetime under the influence of an electromagnetic field and the AC effect
\ie E_{n_s,m_l}=\frac{\mu B}{\eta}\pm\sqrt{m_{0}^{2}+4m_{0}{\vert\tilde{\omega}\vert}\left[n_s+\frac{\vert\tilde{\gamma}_s \vert}{2\eta}-\frac{\tilde{\gamma}_s}{2\eta}\right]},
\label{spectrum2}\fe
where $n_s=n+\frac{1-s}{2}$ is a quantum number, the $+$ sign corresponds to the positive energies states and the $-$ sign corresponds to the negative energies states, respectively. We can see that the energy spectrum \eqref{spectrum2} besides of depending linearly on the strength of the magnetic field $B$ also depends on the parameter $\eta$ associated to deficit angle, the Aharonov-Casher frequency $\tilde{\omega}_{AC}$ and the  Aharonov-Casher phase $\Phi_{AC}$ with periodicity $\Phi_0=\pm 2\pi$, where we have $E_{n_s,m_l}(\Phi_{AC}\pm 2\pi)=E_{n_s,m_l+1}(\Phi_{AC})$. Although the fields \eqref{field1}, \eqref{field2} and \eqref{field3} are restricted to the axis of symmetry of the cosmic string, a region forbidden to the fermion, we see that the fermion energy has electromagnetic contributions due to the  topological defect, or, conical singularity generated by the cosmic string. 

In comparison with the energy spectrum \eqref{spectrum}, the curvature of the cosmic string spacetime slightly changes the pattern of the  energy spectrum oscillations \eqref{spectrum2}. We note that for $0<\eta<1$, the energy values of the fermion in the cosmic string spacetime are slightly larger than in the Minkowski spacetime obtained in the previous section. However, by taking the limit $\eta\to{1}$, we see that we recover the energy espectrum in the Minkowski spacetime. Besides that, we see that in absence of the external electromagnetic field ($B=\tilde{\omega}_{AC}=0$) with $s=+1$, the energy spectrum \eqref{spectrum2} is reduced to the spectrum of the Ref. \cite{Bakke2013} (with $k=cte=0$).  Already to $\mu=0$, the spectrum \eqref{spectrum2} is reduced to that obtained in Ref. \cite{Carvalho} (with $k=cte=0$). Last but not least, also we verified that for $\mu=0$ and $\eta=1$, the spectrum \eqref{spectrum2} is reduced to the energy spectrum of the usual Dirac oscillator \cite{Villalba,Andrade}.

With respect to the eigenfunctions, the original Dirac spinor can be written as
\ie \Psi_{n,m_l}(t,\rho,\theta)=e^{-iEt}\left(
           \begin{array}{c}
            \check{C}_{+}\rho^{\frac{\vert \tilde{\gamma}_{+}\vert}{\eta}}e^{i\left(m_{l}-\frac{1}{2}\right)\theta}e^{-\frac{m_0\vert\tilde{\omega}\vert\rho^2}{2}}F_1\left(-n,\frac{\vert \tilde{\gamma}_{+}\vert}{\eta}+1, m_0\vert\tilde{\omega}\vert\rho^2\right) \\
             i\check{C}_{-}\rho^{\frac{\vert \tilde{\gamma}_{-}\vert}{\eta}}e^{i\left(m_{l}+\frac{1}{2}\right)\theta}e^{-\frac{m_0\vert\tilde{\omega}\vert\rho^2}{2}}F_1\left(-n,\frac{\vert \tilde{\gamma}_{-}\vert}{\eta}+1, m_0\vert\tilde{\omega}\vert\rho^2\right) \\
           \end{array}
         \right)
\label{spinor6},\fe where the new normalization constant is given by
\ie \check{C}_{s}\equiv\frac{\tilde{C}_s(m_0\vert\tilde{\omega}\vert)^{\frac{\vert\tilde{\gamma}_s\vert}{2\eta}}}{\sqrt{2\pi}}
\label{constant}.\fe

Notice that by taking the limit $\eta\to{1}$, we recover the Dirac spinor in the Minkowski spacetime obtained in the previous section.

%-------------------------------------------------------------------
\subsection{Nonrelativistic limite \label{sec4}}

Finally, we analyze the nonrelativistic limit of the DO in the cosmic string spacetime for weak electric and magnetic fields. Let us write $E=\bar{E}+m_0$, where $\bar{E}\ll m_0$, such that the equation \eqref{string9} assumes the form
\ie \left[-\frac{1}{2m_0}\left(\frac{\partial^{2}}{\partial\rho^{2}}+\frac{1}{\rho}\frac{\partial}{\partial\rho}-\frac{\tilde{\Gamma}_s^{2}}{\eta^2\rho^{2}}\right)+\frac{m_{0}\tilde{\omega}^{2}\rho^{2}}{2}-\frac{s\tilde{\omega}}{2}-\frac{2s\tilde{\omega}}{\eta}{\bf S}\cdot{\bf L}-\frac{s\tilde{\omega}\Phi_{AC}}{\pi\eta}+\frac{s\boldsymbol{\mu}\cdot{\bf B}}{\eta}\right]\tilde{\phi}_{s}=\bar{E}\tilde{\phi}_{s}
\label{limite1}, \fe
where
\ie \tilde{\Gamma}_s\equiv L_z+\frac{s\Phi_{AC}}{\pi}, \ {\bf S}=\frac{1}{2}\boldsymbol{\sigma}, \ \boldsymbol{\mu}=\mu\boldsymbol{\sigma}, \ L_z=-i\frac{\partial}{\partial\theta} 
\label{limite2}, \fe
being $\tilde{\phi}_{s}(\rho,\theta)$ given by $\tilde{\phi}_{s}(\rho,\theta)=\frac{e^{il\theta}}{2\pi}f_s(\rho)$, where $f_s(\rho)=(f_+,f_-)^T$ and $l=0,\pm 1, \pm 2,\ldots$.

We verify that the first two terms of Eq. \eqref{limite1} represent the Schr\"odinger-Pauli equation in the cosmic string spacetime for the circular harmonic oscillator under the influence of a electromagnetic field and the AC effect, explaining why this system is called DO \cite{Moshinsky}. The third term of this equation is a constant which sharply shifts all energy levels. The fourth term is a spin-orbit coupling of strength $\tilde{\omega}/\hbar$ (restoring the factor $\hbar$). We see that for the particular case where $\mu=0$ (absence of the electromagnetic field and the AC effect), the Eq. \eqref{limite1} is reduced to the  harmonic oscillator in the cosmic spacetime with a strong spin-orbit coupling \cite{Furtado}. Besides that, by taking the limit $\eta\to{1}$ (with $\mu=0$), the Eq. \eqref{limite1} is reduced to the harmonic oscillator in Euclidean space without curvature \cite{Andrade}.

Now, using the prescription $E=\bar{E}+m_0$ ($\bar{E}\ll m_0$) into energy spectrum \eqref{spectrum2}, we obtain  
\ie \bar{E}_{n_s,m_l}=\frac{\mu B}{\eta}+2{\vert\tilde{\omega}\vert}\left[n_s+\frac{\vert\tilde{\gamma}_s \vert}{2\eta}-\frac{\tilde{\gamma}_s}{2\eta}\right].
\label{spectrum3}\fe

We verify that by making $\mu=0$, the spectrum \eqref{spectrum3} is reduced to the spectrum of the harmonic oscillator in cosmic string spacetime with a strong spin-orbit coupling \cite{Furtado}. Now, taking the limit $\eta\to{1}$ (with $\mu=0$), the spectrum \eqref{spectrum3} is reduced to spectrum of harmonic oscillator in Euclidean space without curvature with a strong spin-orbit coupling \cite{Andrade}, confirming the consistency of our results.

%-------------------------------------------------------------------------
\section{Conclusion\label{conclusion}}

In this paper, we study the $(2+1)$-dimensional DO in the presence of an electromagnetic field and the Aharonov-Casher effect in the Minkowski spacetime and the cosmic string spacetime. In both scenarios, we verify that the eigenfunctions of the DO are written in terms of the confluent hypergeometry functions.  Also, the energy spectrum was determined exactly, and we can observe a dependence on the strength of the magnetic field $B$, the Aharonov-Casher frequency $\omega_ {AC}$, and the Aharonov-Casher phase $\Phi_{AC}$. Besides that, in the Minkowski spacetime, We noticed that the DO could be set to resonate perfectly with the half-frequency of $\omega_ {AC}$, such that the total energy of the particle and antiparticle is a physics constant that depends on its mass of rest and the energy acquired by the magnetic field. In relation to the cosmic string spacetime, we observed that for $0<\eta<1$, where $0<\eta\leq 1$ is a parameter related to the deficit angle of the system, the energy values are larger than in the Minkowski spacetime. Besides that, in the limit $\eta\to{1}$, we recover all results in the Minkowski spacetime. Moreover, we see that in the absence of the external electromagnetic field, the energy spectrum (curved and flat spacetime) is reduced to the spectrum of the Ref. \cite{Bakke2013}. Already to $\mu=0$ (and after $\eta=1$), the energy spectrum (curved and flat spacetime) is reduced to the spectrum of Refs. \cite{Carvalho,Villalba,Andrade}. We finish this work with the nonrelativistic limit of our results in the cosmic string spacetime. We observed that in this limit, we reduced the DO to the circular quantum harmonic oscillator with a strong spin-orbit coupling. Already for $\mu=\eta=1$, the nonrelativistic energy spectrum is reduced to the spectrum of the harmonic oscillator in Euclidean space without curvature \cite{Andrade}.

%-------------------------------------------------------------------------
\section*{Acknowledgments}
\hspace{0.5cm}The authors would like to thank the Funda\c{c}\~{a}o Cearense de Apoio ao Desenvolvimento Cient\'{\i}fico e Tecnol\'{o}gico (FUNCAP)(PNE-0112-00061.01.00/16), and the Conselho Nacional de Desenvolvimento Cient\'{\i}fico e Tecnol\'{o}gico (CNPq) for grants Nos. 305678/2015-9 (RVM), and 308638/2015-8 (CASA).

\end{document}